\documentstyle[aps,preprint]{revtex}

\begin{document}


\preprint{YITP-96-51}
\title{
D-brane configuration and black hole thermodynamics
}
\author{
Shinji Mukohyama
}
\address{
Yukawa Institute for Theoretical Physics, Kyoto University \\ 
Kyoto 606-01, Japan
}
\date{\today}
\maketitle


\begin{abstract}
We consider a configuration of strings and solitons in the type IIB
superstring theory on $M^5\times T^5$, which is composed of a set of
arbitrarily-wound D-fivebranes on $T^5$ and a set of arbitrarily-wound
D-strings on $S^1$ of the torus. For the configuration, it is shown
that number of microscopic states is bounded from above by the exponential
of the Hawking-Bekenstein entropy of the corresponding black hole and
the temperature of closed string radiation from the D-branes is
bounded from below by the Hawking temperature of the black hole. After
discussing the necessary and sufficient condition to saturate these
bounds, we give some speculations about black hole thermodynamics.
\end{abstract}

\vfill



\section{Introduction}
Black hole thermodynamics \cite{Bekenstein} \cite{Hawking} is one of
the most interesting topics in the theory of black holes. Recently
Strominger and Vafa \cite{Strominger&Vafa} gave a microscopic origin
of the Hawking-Bekenstein entropy of a black hole in the framework of
superstring theory by using the D-brane technology
\cite{Polchinski}. Hawking radiation was also explained in the D-brane
picture \cite{Callan&Maldacena}.

In this letter we consider a configuration of strings and solitons in
the type IIB superstring theory on $M^5\times T^5$.  The corresponding
classical black-brane solution was obtained in \cite{HMS} and is
characterized by four charges ($Q_1$, $Q_5$, $n_L$, $n_R$). In the
case of the extreme limit ($n_R=0$), Callan and Maldacena have thought of
it as a configuration composed of $Q_5$ D-fivebranes wrapped on the
torus and $Q_1$ D-strings wrapped on $S^1$ of the torus, based on the
fact that a D-string has charge ($Q_1$, $Q_5$)$=$($1$, $0$) and a
D-fivebrane has charge ($Q_1$, $Q_5$)$=$($0$, $1$). On the contrary
Maldacena and Susskind \cite{Maldacena&Susskind} regarded it as a
configuration of a single long D-string and a single long
D-fivebrane. In both interpretations $P_L\equiv n_L/R$ is the momentum
along the circle which is a sum over left moving massless modes of
open strings on the D-branes, where $R$ is the radius of the
circle. In this letter we consider a more general configuration of
D-branes and investigate open strings on it.

In Sec. \ref{sec:number} we give the number of microscopic states, which
is related to the black hole entropy. In Sec. \ref{sec:canonical}
temperature of a canonical ensemble of open strings on D-branes is
obtained. It is related to the temperature of a decay of D-brane
excitations to a closed-string mode, which seems to be the Hawking
radiation. Finally in Sec. \ref{sec:summary} we summarize this letter
and give some speculations. 


\section{Number of microscopic states}	\label{sec:number}
First let us consider a set of $Q_1$ D-strings wrapped on $S^1$ whose
length is $2\pi R$. They may connect up \cite{Maldacena&Susskind} to
form a set of multiply-wound D-strings, which is composed of
$N_{q_1}^{(1)}$ D-strings of length $2\pi Rq_1$
($q_1=1,2,\cdots$). Next consider a set of $Q_5$ D-fivebranes wrapped on
$T^5$ and a set of $Q_1$ D-strings wrapped on $S^1$ of the torus. The
corresponding non-extremal supergravity solution is obtained by
setting $N_{\bar{1}}=N_{\bar{5}}=0$ in the solution found in
\cite{HMS} and is characterized by four parameters ($Q_1$, $Q_5$,
$n_L$, $n_R$). Here $Q_1$ and $Q_5$ are charges of a U(1) gauge field;
$n_L$ ($n_R$) is left (right) moving momentum along $S_1$ multiplied
by $R$. By suppressing reference to the other four directions we may
think of the D-fivebranes as D-strings wrapped on the circle
\cite{Maldacena&Susskind}. The D-fivebranes may also connect up
to form a set of multiply-wound D-fivebranes, which is composed of
$N_{q_5}^{(5)}$ D-fivebranes of length $2\pi Rq_5$ ($q_5=1,2,\cdots$)
along the circle. Note that $N_{q_1}^{(1)}$ and $N_{q_5}^{(5)}$ must
satisfy the following constraints: 
\begin{eqnarray}
 Q_1 & = & \sum_{q_1} q_1N_{q_1}^{(1)},	\nonumber	\\
 Q_5 & = & \sum_{q_5} q_5N_{q_5}^{(5)}	\nonumber
\end{eqnarray}
since a D-string which winds $q_1$ times has charge ($Q_1$,
$Q_5$)$=$($q_1$, $0$) and a D-fivebrane which winds $q_5$ times has
charge ($Q_1$, $Q_5$)$=$($0$, $q_5$).

There are several types of open strings on the D-brane configuration:
both boundaries are on a common D-string; one boundary is on a
D-string and another is on a different D-string; one is on a D-string
and another is on a D-fivebrane; etc. Among them, as an approximation, 
we consider only strings which connect a $q_1$-wound D-string and a
$q_5$-wound D-fivebrane and neglect contribution from other
strings. The spectrum of the strings is 
\begin{eqnarray}
 p_n(q_1,q_5) & = & \frac{\pm n}{R(q_1,q_5)_{LCM}}, 
	\quad (n=1,2,\cdots )	\label{eqn:spectrum}\\
 ('+' \mbox{ for left moving} & ; & '-' \mbox{ for right moving}),
			\nonumber 
\end{eqnarray}
where $(q_1,q_5)_{LCM}$ is the least common multiple of $q_1$ and
$q_5$, since the boundary condition of the strings is 
\begin{equation}
 X^5 \sim X^5 + 2\pi R (q_1,q_5)_{LCM}.	\nonumber
\end{equation}

We also regard the momentum $P_L\equiv n_L/R$ ($P_R\equiv -n_R/R$)
along the circle as a sum over the left (right) moving massless modes
of the strings. Hence the number of microscopic states $d(n_L,n_R)$ is 
given by
\begin{eqnarray}
 d(n_L,n_R) & = & d(n_L)d(n_R),	\nonumber	\\
 \sum_N d(N)w^N & = & \prod_{q_1}\prod_{q_5}\prod_{n=1}^{\infty}
	\left[\frac{1+w^{n/(q_1,q_5)_{LCM}}}{1-w^{n/(q_1,q_5)_{LCM}}}
	\right]^{4N_{q_1}^{(1)}N_{q_5}^{(5)}}.	\nonumber
\end{eqnarray}
By using the asymptotic form of the $\theta$-function and the
saddle point method, we can obtain $d(n_{L,R})$ in the limit 
\begin{equation}
 \sqrt{\frac{Q_1Q_5}{n_{L,R}}}\ll \min_{(q_1,q_5)\in A}(q_1,q_5)_{LCM},
	\label{eqn:limit}
\end{equation}
where $A \equiv \{ (q_1,q_5)|N_{q_1}^{(1)}N_{q_5}^{(5)}\neq 0\}$. The
result is
\begin{eqnarray}
 d(n_{L,R}) & \approx & \exp\left[ 2\pi\sqrt{n_{L,R}\sum_{q_1}\sum_{q_5}
	N_{q_1}^{(1)}N_{q_5}^{(5)}(q_1,q_5)_{LCM}}\right]
						\nonumber	\\
 & \leq & \exp (2\pi\sqrt{n_{L,R}Q_1Q_5}).	\nonumber
\end{eqnarray}
Therefore, for a configuration satisfying (\ref{eqn:limit}),
$d(n_L,n_R)$ is bounded from above: 
\begin{equation}
 d(n_L,n_R) \leq \exp\left[ 
	2\pi\left(\sqrt{n_L}+\sqrt{n_R}\right)\sqrt{Q_1Q_5}\right],
						\nonumber
\end{equation}
where the bound is saturated if and only if all $(q_1,q_5)$ $(\in A)$
are relatively prime.


\section{Canonical ensemble of open strings}	\label{sec:canonical}
In the D-brane approach to Hawking radiation \cite{Callan&Maldacena}
\cite{Das&Mathur} \cite{Maldacena&Strominger}, the thermal 
spectrum of the decay of non-BPS D-brane excitations to closed string
modes was obtained by summing up decay rates over all consistent
initial states of open strings on the D-branes. The summation was
approximately performed by replacing the microcanonical ensemble by a
canonical ensemble. The resulting temperature of the decay spectrum is
as follows \cite{Maldacena&Strominger}:
\begin{equation}
 T_{decay} = T,
\end{equation}
where $T$ is the temperature of the canonical ensemble.

Therefore, in this section we consider a canonical ensemble of open
strings on the D-brane configurations introduced in
Sec. \ref{sec:number}. The partition function is
\begin{eqnarray}
 Z(\beta,\alpha ) & = & Z_L(\beta,\alpha )Z_R(\beta,\alpha ),
					\nonumber\\
 Z_{L,R}(\beta,\alpha ) & = & 
	\prod_{q_1}\prod_{q_5}\prod_{n=1}^{\infty}
	\left[\frac{1+e^{-\beta e_n(q_1,q_5)-\alpha p_n(q_1,q_5)}}
	{1-e^{-\beta e_n(q_1,q_5)-\alpha p_n(q_1,q_5)}}
	\right]^{4N_{q_1}^{(1)}N_{q_5}^{(5)}},	\nonumber
\end{eqnarray}
where $p_n$ is given by (\ref{eqn:spectrum}) and $e_n=|p_n|$. By using
the asymptotic form of the $\theta$-function, we can obtain 
$Z_{L,R}(\beta,\alpha )$ in the limit (\ref{eqn:limit}). The result is 
\begin{eqnarray}
 \ln Z_{L,R}(\beta,\alpha ) &  = & \frac{\pi^2R}{\beta\pm\alpha}
	\sum_{q_1}\sum_{q_5}N_{q_1}^{(1)}N_{q_5}^{(5)}(q_1,q_5)_{LCM},
						\nonumber	\\
 ('+' \mbox{ for } 'L' & ; & '-' \mbox{ for }'R').	\nonumber
\end{eqnarray}
$\beta$ and $\alpha$ are fixed by 
\begin{eqnarray}
 E_L & = & P_L = n_L/R,	\nonumber	\\
 E_R & = & -P_R = n_R/R,	\nonumber
\end{eqnarray}
where $E_L$ ($E_R$) is the left (right) movers' contribution to the
expectation value of the energy and $P_L$ ($P_R$) is the left (right)
movers' contribution to the expectation value of the momentum:
\begin{eqnarray}
 E_L+E_R & = & -\frac{\partial\ln Z(\beta,\alpha )}{\partial\beta},
			\nonumber	\\
 P_L+P_R & = & -\frac{\partial\ln Z(\beta,\alpha )}{\partial\alpha}.
			\nonumber
\end{eqnarray}
Thus, for a configuration satisfying (\ref{eqn:limit}), the entropy
$S$ and the inverse-temperature $1/T=\beta$ are
\begin{eqnarray}
 S & = & 2\pi\left(\sqrt{n_L}+\sqrt{n_R}\right)
	\sqrt{\sum_{q_1}\sum_{q_5}
	N_{q_1}^{(1)}N_{q_5}^{(5)}(q_1,q_5)_{LCM}},	\nonumber\\
 \frac{1}{T} & = & \frac{\pi R}{2}
	\left(\frac{1}{\sqrt{n_L}}+\frac{1}{\sqrt{n_R}}\right) 
	\sqrt{\sum_{q_1}\sum_{q_5}
	N_{q_1}^{(1)}N_{q_5}^{(5)}(q_1,q_5)_{LCM}}.	\nonumber
\end{eqnarray}
They are bounded from above as follows:
\begin{eqnarray}
 S & \leq & 2\pi\left(\sqrt{n_L}+\sqrt{n_R}\right)\sqrt{Q_1Q_5},\\
 \frac{1}{T} & \leq & \frac{\pi R}{2}
	\left(\frac{1}{\sqrt{n_L}}+\frac{1}{\sqrt{n_R}}\right)
	\sqrt{Q_1Q_5},
\end{eqnarray}
where the the bounds are saturated if and only if all $(q_1,q_5)$
$(\in A)$ are relatively prime. Finally note that the following
relation is trivially satisfied:
\begin{equation}
 S \approx \ln d(n_L,n_R).	\nonumber
\end{equation}


\section{Summary and speculations}	\label{sec:summary}
In summary we have investigated open strings on the general D-brane
configurations. For configurations satisfying (\ref{eqn:limit}), the
number of microscopic states $d(n_L,n_R)$ is bounded from above and
the temperature of a decay of D-brane excitations to closed strings is
bounded from below. Moreover, for any configurations not satisfying
(\ref{eqn:limit}), the same bounds seem to exist without saturation
\cite{Maldacena&Susskind}. Therefore it is expected that the
quantities calculated from the microscopic point of view are related
as follows to the macroscopic quantities of the black hole.
\begin{eqnarray}
 \ln d(n_L,n_R) & \leq & S_{BH},	\label{eqn:lndN}\\
 T_{decay} & \geq & T_{BH},	\label{eqn:Tdecay}
\end{eqnarray}
where $S_{BH}$ and $T_{BH}$ are the Hawking-Bekenstein entropy and
the Hawking temperature of the black hole \cite{HMS}
\cite{Maldacena&Strominger}:
\begin{eqnarray}
 S_{BH} & = & 2\pi\left(\sqrt{n_L}+\sqrt{n_R}\right)\sqrt{Q_1Q_5},
			\nonumber\\
 \frac{1}{T_{BH}} & = & \frac{\pi R}{2}
	\left(\frac{1}{\sqrt{n_L}}+\frac{1}{\sqrt{n_R}}\right)
	\sqrt{Q_1Q_5}.	\nonumber
\end{eqnarray}
Note that the Hawking-Bekenstein entropy and the Hawking temperature
are defined by macroscopic quantities (area and surface gravity of the 
event horizon). The bounds (\ref{eqn:lndN}) and (\ref{eqn:Tdecay}) are
saturated if and only if all $(q_1,q_5)$ $(\in A)$ are relatively
prime and (\ref{eqn:limit}) is satisfied. 

Thus several speculations may be possible.
\begin{itemize}
 \item
Inside a black hole characterized by the four parameters ($Q_1$,
$Q_5$, $n_L$, $n_R$), some dynamical processes may occur. The
processes may be described in the D-brane picture: D-branes
repeat fission and fusion to settle down to one of the states for
which all $(q_1,q_5)$ $(\in A)$ are relatively prime and
(\ref{eqn:limit}) is satisfied.
 \item 
During the process the microscopic entropy increases to reach the
Hawking-Bekenstein entropy of the corresponding black hole. Moreover
the temperature of closed string radiation from the D-branes decreases
to reach the Hawking temperature of the black hole. 
\end{itemize}

These speculations may be significant to investigate the microstates
of dynamical black holes. For example let us consider a merger of two
black holes $B_1$ and $B_2$ and suppose that $B_1$ corresponds to a
D-brane configuration $\{ N_{q_1}^{(11)}, N_{q_5}^{(15)}\}$ and $B_2$
corresponds to a configuration 
$\{ N_{q_1}^{(21)}, N_{q_5}^{(25)}\}$. Just after merging, the large
black hole $B$ formed by the merger corresponds to a configuration 
$\{ N_{q_1}^{(1)}=N_{q_1}^{(11)}+N_{q_1}^{(21)}, 
N_{q_5}^{(5)}=N_{q_5}^{(15)}+N_{q_5}^{(25)} \}$, provided that
directions of the D-strings are the same for $B_1$ and $B_2$. In general
the last configuration does not saturate the bounds (\ref{eqn:lndN})
and (\ref{eqn:Tdecay}) even when the configurations for $B_1$ and
$B_2$ saturate the bounds. Thus, in general just after the merger the
microscopic entropy of $B$ does not agree with the corresponding
Hawking-Bekenstein entropy and the temperature of the closed string
radiation does not agree with the corresponding Hawking
temperature. However, after a sufficiently long time the D-branes'
fission and fusion settle the entropy and the temperature to the
Hawking-Bekenstein entropy and the Hawking temperature. To say
something quantitative about the above process, we have to find a
supergravity solution which represents the merger of branes and have to
investigate the D-branes' fission and fusion in terms of the
elementary interactions. However, without doing these, we could reach
the above conclusion about the dynamical process inside the black hole 
by analyzing the microscopic entropy of open strings on D-branes.

\vspace{3cm}
\centerline{\bf Acknowledgment}
The author gratefully thanks H. Kodama for continuous encouragement
and critical discussions. He also thanks H. Ishikawa, K. Sugiyama and
Y. Matsuo for helpful discussions. He is grateful to S. A. Hayward for 
his careful reading of the English.


\end{document}